# Bose-Einstein Condensation in Financial Systems

**Kestutis Staliunas**

*PTB Braunschweig, Bundesallee 100, 38116 Braunschweig, Germany*

*kestutis.staliunas@ptb.de*

**Abstract**

We describe financial systems as condensates, similar to Bose-Einstein condensates, and calculate statistical distributions following from the model. The calculated distributions of investments into speculated financial assets are found equivalent to a Pareto distribution, and the calculated distributions of the price moves are found equivalent to exponentially truncated Levy distributions.

One indication that financial markets are related with Bose-Einstein condensates (BECs) is the similarity of statistical distributions in both systems. It is generally accepted that the statistical distributions of variations of prices of financial assets (shares, indices, commodities, exchange rates) are power laws [1], and (exponentially) truncated power laws [2]. The Bose-Einstein distribution for the occupation of energy states is also an exponentially truncated power law: the average occupation of energy states in BECs $n(E) = 1/(\exp((E-\pmb{m})/kT)-1)$ shows the asymptotics: $n(E) \propto (E/kT)^{-1}$ for small energies (for the condensed part of BEC) and $n(E) \propto \exp(-E/kT)$ for large energies (noncondensed part). The power law exponents in finance markets are however different from those in BECs: the distribution of wealth follows the Pareto law: $n(w) \propto w^{-1-\pmb{a}}$ [3], where the Pareto exponent in empirical studies is found in region $1 < \pmb{a} < 2$ [4]. The distributions of price moves for most financial assets follow a truncated Levy distribution [2], with the asymptotics for small and moderate price variations $\Delta x$ given by a power law: $p(\Delta x) \propto \Delta x^{-1-\pmb{g}}$, and for large variations given by an exponential tail: $p(\Delta x) \propto \exp(-\Delta x)$. The Levy exponent is empirically found in range $1.3 < \pmb{g} < 1.8$ [5]. Despite of the significant differences in power law exponents (the power law exponents for Bose-Einstein distributions are $\pmb{a} = \pmb{g} = 0$), the fact that the distributions both in financial markets and BECs follow an exponentially truncated power law is remarkable.

Another indication that the financial markets are related with BECs is that both systems are partially random, and partially coherent. The atom collisions in classical gases are completely random (constrains being just the energy and momentum conservation), which leads to Maxwell-Boltzman distributions. The particle collisions in bosonic gases are selective, in that the atoms after collisions prefer to choose occupied states, due to the bosonic enhancement effect. Evidently processes in finance market are also on one hand chaotic and unpredictable, like chaotic collisions of atoms in classical gases. On the other hand the events in finance markets are somehow motivated. The motivation in general brings order and coherence into a system. This simultaneous presence of randomness and of coherence hints on deeper relations between the finance and BEC systems.

A common physics in BECs and finance systems bases on a similar mechanism of the coherence in both systems. As noted above, the bosonic enhancement is responsible for the coherence in atomic (or photonic) condensates, in that the quantum particles tend to choose occupied states. In finance, one obvious behavior scenario is that most market participants tend to invest like the others participants, i.e. to occupy more "attractive", more "popular", in general already occupied, states. This is due to a choice of investing strategies according to the opinion of majorities. This is also due to a "condensation" of investors into investment groups, with common investment strategies. In overall the so called herding effect in economy and finance is plausible [6]. Evidently the finance markets are somewhat more complicated than bosonic gas, and other motivations than herding play a role here. For instance every market participant is motivated to maximize his wealth, i.e. to optimize the outcomes of his financial deals.

One of first models for finance markets, that of Bachelier [7] compares the stochastic diffusion of the market prices with stochastic diffusion of a Brownian particle. The Brownian particle is in a thermal equilibrium with the atoms of the environment, like the price is in an equilibrium with the kinetics of the market participants. The Bachelier model leads to Gaussian distributions for price moves, in analogy with the Maxwell distribution of atom velocities in classical gases. However, if one draws an analogy between finance markets and partially coherent gases, then the price, being in thermal equilibrium with Bose particles would not obey Gaussian distributions, but rather the Bose-Einstein distributions, i.e. would show the power laws.

In this letter we substantiate the idea that the finance markets are analogous to Bose gases. We consider the two motivations discussed above for the behaviour of market participants: 1) herding (also present in Bose gases); 2) optimization of the outcomes of deals (absent in Bose gases). Apart from these two global motivations we consider the financial deals as completely stochastic (or driven by a variety of different individual motivations, not possible to be incorporated in a macroscopic description). We derive statistical distributions of finance indices based on these assumptions. The derived distributions correspond well with the distributions observed in finance markets, i.e. with the Pareto distributions of wealth, and with the exponentially truncated Levy distributions of price moves.

We simplify maximally the model by assuming that each ($i$-th) market participant occupies states in a two dimensional space $X_i = (m_i, s_i)$, where $m_i$ is the amount of "money", and $s_i$ is the amount of "shares" in possession. In general "money" $m$ is some exogeneous asset, in the sense that the investors can buy from it as much as they wish, and the "stock" $s$ is some "risky" asset subjected to speculation. The generalization to the systems of many sorts of shares is possible. The considered one share system is illustrated in Fig.1. An elementary deal involving two market participants means a buying or selling of a particular amount of shares, i.e. a change of the states of participants, as indicated by arrows. We assume "two particle collisions" only, i.e. the deals between two participants. In reality more than two participants may participate in deals, however, without losing generality we can decompose complicated deals into two participant elementary deals.

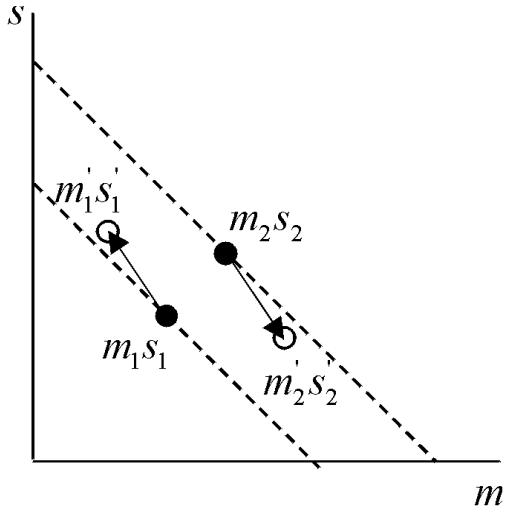

Fig.1. Phase space of "one share" system. Market participants occupy the states in the phase space parameterized by "money" $m$, and "shares" $s$. A deal between to two market participants $X_1 = (m_1, s_1)$ and $X_2(m_2, s_2)$ corresponds to a jump to new states $X_1' = (m_1', s_1')$ and $X_2'(m_2', s_2')$ conserving the total amount of money and shares.

Each deal conserves the total amount of money and of shares possessed by both participants of the deal. (The arrows in Fig.1 are directed oppositely, and are of equal length.) The direction of the arrows indicates an agreed price of the share for a particular deal. If every deal would occur at a fixed price, the individual wealth of investors $r_i = m_i + s_i$, would not vary in time, and no thermalisation and equipartition in the ensemble would occur. However, some deals can be profitable for one participant, and brings losses to another one (in Fig.1 the deal is profitable for participant 1), also the prices fluctuate, therefore mixing in the system occurs, and the system should reach a thermal equilibrium.

We apply a textbook technique [8], to calculate the average occupations of the states in the parameter space: Assuming that two particles (two market participants) involved in a collision (a deal) were initially in states $X_1 = (m_1, s_1)$ and $X_2(m_2, s_2)$, with the average occupations $n_1$ and $n_2$, and that after collision they occupy new states $X_1' = (m_1', s_1')$ and $X_2'(m_2', s_2')$, with average occupations $n_1'$ and $n_2'$, the probability of the above collision is: $n_1 n_2 (1 + n_1')(1 + n_2')$. Here the probability of a particular collision is proportional to the occupation of initial states, since the colliding particles (the deal partners) must meet one another, and depends on the occupation of the final states, due to Bosonic enhancement (herding) effect. A detailed balance requires that the probability of the transition in the reverse direction is equal to that of the forward transition, i.e. $n_1 n_2 (1 + n_1')(1 + n_2') = n_1' n_2' (1 + n_1)(1 + n_2)$, which can be rewritten:

$$\frac{n_1}{(1+n_1)}\frac{n_2}{(1+n_2)} = \frac{n_1^{'}}{(1+n_1^{'})}\frac{n_2^{'}}{(1+n_2^{'})}. \quad (1)$$

The solution of (1) taking into account conserved quantities leads to:

$$\frac{n(m,s)}{1+n(m,s)} = \exp[\boldsymbol{b}(\boldsymbol{m}-m-s)], \quad (2)$$

with $\boldsymbol{m}$ having the meaning of a chemical potential, and indicating the level of condensation in the system, and $\boldsymbol{b} = 1/(kT)$ having the meaning of inverse temperature. (2) easily leads to cellebrated Bose-Einstein $n(m,s) = (\exp[\boldsymbol{b}(-\boldsymbol{m}+m+s)]-1)^{-1}$.

We modify (1), (2) for application in financial markets. First, we assume that the herding effects only the risky asset $s$, but not money $m$. It would be unrealistic to assume that a market participants finds the state with less money more attractive because the majority is poorer than he, however it is realistic to assume that the market participants would sell shares if everybody else were selling. With this assumption the attractivity of a state should not be $(1+n(m,s))$ as in BECs, but rather $(1+n(s))$, where $n(s) = \int n(m,s)dm$ is the distribution in $s$ space, regardless of money $m$. Next, the attractivity for the state in financial markets is evidently proportional to its wealth $r = m+s$, therefore the probability of the jump from the state $X = (m,s)$ to the state $X^{'} = (m^{'},s^{'})$ is proportional to $r^{'}/r = (m^{'}+s^{'})/(m+s)$. With this in mind the analog of (2) now reads:

$$\frac{n(m,s)}{1+n(s)} = (s+m)^2 \exp[\boldsymbol{b}(\boldsymbol{m}-m-s)]. \quad (3)$$

Integration of (3) with respect to $m$ allows to calculate the distribution in $s$ space:

$$n(s) = \frac{1+\boldsymbol{b}s+(\boldsymbol{b}s)^2/2}{\exp[\boldsymbol{b}(\Delta\boldsymbol{m}+s)]-(1+\boldsymbol{b}s+(\boldsymbol{b}s)^2/2)}, \quad (4)$$

here $\Delta\boldsymbol{m} = \boldsymbol{m}_0 - \boldsymbol{m}$ is the normalized chemical potential: $\boldsymbol{m}_0 = \ln(2/\boldsymbol{b}^3)/\boldsymbol{b}$. (3) and (4) allow to calculate the full distribution:

$$n(m,s) = \frac{(s+m)^2 \exp[\boldsymbol{b}(\boldsymbol{m}_0-m)]}{\exp[\boldsymbol{b}(\Delta\boldsymbol{m}+s)]-(1+\boldsymbol{b}s+(\boldsymbol{b}s)^2/2)}. \quad (5)$$

(4) and (5) are central distributions as following from our condensate model.

(5) indicates, that the condensation occurs in the space of the speculated asset $s$. In $m$ space no condensation occurs, and the distributions are Poisson - like: for not condensed markets $\boldsymbol{b}\Delta\boldsymbol{m} \gg 1$: $n(m) = \int n(m,s)dm \propto (1+\boldsymbol{b}m+(\boldsymbol{b}m)^2/2)\exp(-\boldsymbol{b}m)$ has the maximum at zero; in the limit of strong condensation $b\Delta m \ll 1$: $n(m) \propto (bm)^2 \exp(-bm)$, has the maximum at $bm_0 = 2$. The distribution in $s$ space (4) leads to the following asymptotics: for highly condensed markets (4) leads to $n(s) = \left(b\Delta m + (bs)^3/6\right)^{-1}$, which saturates to $n_0 = (b\Delta m)^{-1}$ for $bs \to 0$, and results in a Pareto wealth distribution $n(s) = s^{-1-a}$ with the power exponent $a = 2$. The not condensed markets $b\Delta m \gg 1$, and/or not condensed tails of condensed markets $bs \gg 1$ obey an exponential law $n(s) = \exp[-b(\Delta m + s)]$. The distributions for the financial systems with different condensation degrees are plotted in Fig.2.a. One generally obtains 1) plateau (saturation) for small values of $s$; 2) power law region for intermediate values of $s$. The power law region increases with increasing condensation degree; 3) exponential decay for large values of $s$.

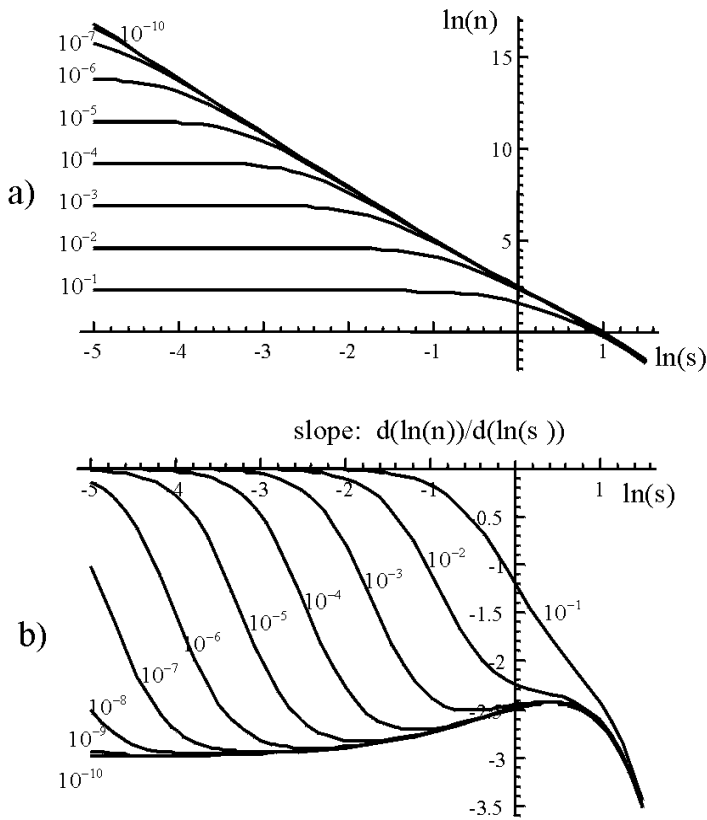

Fig.2. Average occupations of share states in double logarithmic representation a), and their local slopes (local Pareto exponents) b) for different values of normalized chemical potential $\Delta m$, as obtained from (4). $b = 1$.

Fig.2.b. shows the local slopes of the double-logarithmic plot of occupation distribution. At a condensation threshold a region of power law with Pareto exponent $a \approx 1.4$ emerges. With the increasing condensation level the Pareto exponent increases, up to a limiting value $a = 2$ for perfectly condensed markets.

The relation between the normalized chemical potential $\Delta m$, and the integral quantities, such as total numbers of participants $N = \iint n(m,s)dmds$, of money $M = \iint m \cdot n(m,s)dmds$, and of shares $S = \iint s \cdot n(m,s)dmds$) is not analytic. In a limit of high condensation: $N \propto b^{-5/3}\Delta m^{-2/3}$, $M \propto b^{-8/3}\Delta m^{-2/3}$, and $S \propto b^{-7/3}\Delta m^{-1/3}$. The number of particles is in units of a critical (visible) occupation number, when the state becomes attractive. Recall that the attractivity of the state is $(1+n(s))$. The parameters for condensation threshold depends on its definition: if one defines the condensation threshold as appearance of plateau In Fig.2, then this occurs at $b\Delta m_{thr} \approx 0.0037$, and $bN_{thr} \approx 105$.

Next we derive the distribution of price moves for a traded asset $s$, based on the distribution (5), and assuming, that the price is in equilibrium with the microscopic dynamics of the system, i.e. that the price change is proportional to the difference between the demand and supply. This means that the distribution of price change is proportional to the distribution of the jumps $n(\Delta s)$ in the phase space of the system. Using the above postulated probability for a jump $(m,s) \rightarrow (m',s') = (m-\Delta m, s+\Delta s)$:

$$n(m \rightarrow m', s \rightarrow s') = \frac{m'+s'}{m+s} n(m,s)\left(1 + \int n(m',s')dm'\right). \qquad (6)$$

and integrating with respect to all possible initial states: $\Delta m < m < \infty$ (one must posses at least the amount of money $\Delta m$ to buy a share), and $0 < s < \infty$, and with respect to all $\Delta m$, the distribution of a size of a deal $n(\Delta s)$ is obtained, however, does not leads to analytically tractable results. Even limiting to "fair deals", i.e. to the deals at fixed prices $\Delta s/\Delta m = 1$, the distribution of deal sizes $n(\Delta s)$ is complicated:

$$n(\Delta s) = \int \frac{(1+b(s+\Delta s)+(b(s+\Delta s))^2/2)\exp[b(\Delta m+s)]ds}{(\exp[b(\Delta m+s)]-(1+bs+(bs)^2/2))(\exp[b(\Delta m+s+\Delta s)]-(1+b(s+\Delta s)+(b(s+\Delta s))^2/2))} \qquad (7)$$

The family of distributions as obtained by numerical integration of (7), is given in Fig.3, shows sharply peaked, and exponentially decaying distributions very similar to those found in financial data. The family of distributions in double logarithmic representation is plotted in Fig.4.a. The picture, similar to that in Fig.2 for the distribution of investments, is obtained: 1) the distribution of price changes saturates for small $\Delta s$, 2) the distribution of price changes follows the power law for intermediate $\Delta s$: $n(\Delta s) = \Delta s^{-1-g}$, with Levy exponent $g$ is in the range of $1.3 < g < 1.8$; 3) exponential decay for large $\Delta s$: $n(\Delta s) = b^{-1}\exp[-b(\Delta m+\Delta s)]$.

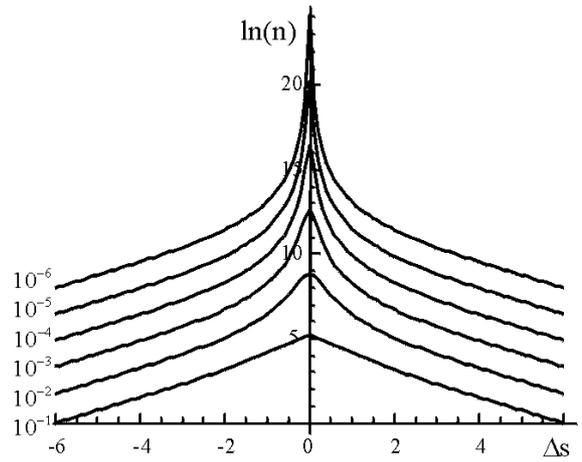

Fig.3. Distribution of the price changes in linear logarithmic representation for different values of normalized chemical potential $\Delta m$, as obtained from numerical integration of (7). $b = 1$.

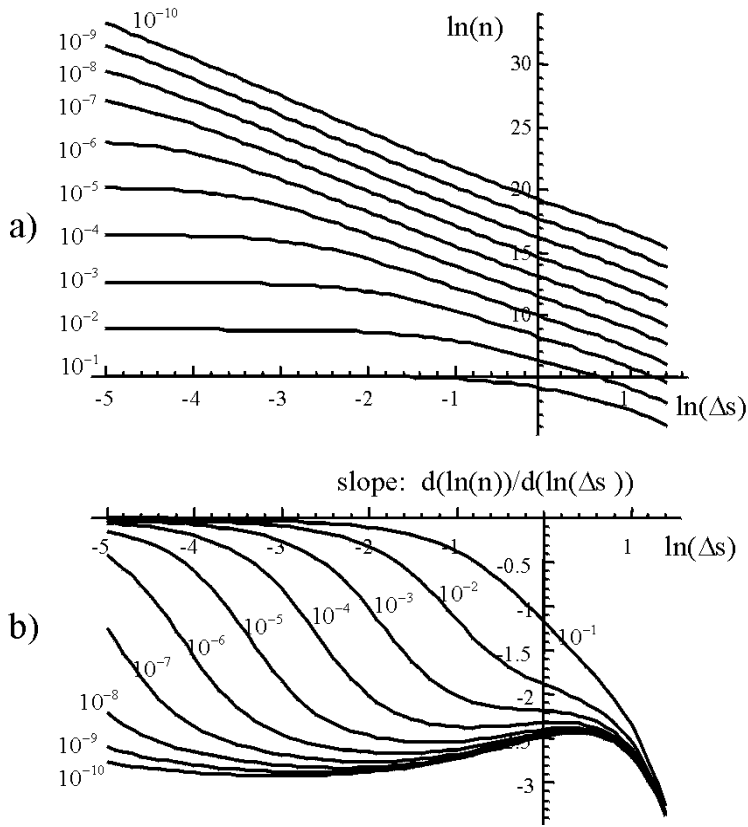

Fig.4. Distribution of price changes in double logarithmic representation a), and the local slopes of the distribution (local Levy exponents) b), for different values of normalized chemical potential $\Delta m$, as obtained from (7). $b = 1$.

Concluding, we consider financial system as a partially random partially coherent bosonic system. We derive statistical distributions based on only two ingredients of the behaviour of market participants: 1) that the individual market participants tend to cluster, and to behave according to the opinion of majority; 2) that the market participants seek for profit. Considering the first ingredient only, a Bose-Einstein distribution is recovered. Although the Bose-Einstein distribution is identical to exponentially truncated Pareto and Levy distributions, the corresponding power law exponents: $a = g = 0$ are significantly different from those observed in financial systems. Accounting for the second ingredient leads to exponentially truncated Pareto and Levy distributions with power exponents corresponding well to the ones observed in financial markets.

These Pareto exponents are found empirically in the limits $1 < a < 2$. The results from our model are compatible with these observations. The empirical Pareto exponents are not very precise, since the statistical data on the wealth distribution could be biased. The empirical Levy exponents for price variation are of better confidence; they are reported mostly in the region $1.3 < g < 1.8$, which corresponds very well to those following from our model. The correspondence between the power law exponents following from our BEC model are also compatible with those recently calculated from kinetic models of finance markets [9].

The work has been supported by Sonderforschungsbereich 407 of Deutsche Forschungsgemeinschaft.